# Codon Context Optimization in Synthetic Gene Design

Dimitris Papamichail, Hongmei Liu, Vitor Machado, Nathan Gould, J. Robert Coleman, Georgios Papamichail

**Abstract**—Advances in de novo synthesis of DNA and computational gene design methods make possible the customization of genes by direct manipulation of features such as codon bias and mRNA secondary structure. Codon context is another feature significantly affecting mRNA translational efficiency, but existing methods and tools for evaluating and designing novel optimized protein coding sequences utilize untested heuristics and do not provide quantifiable guarantees on design quality. In this study we examine statistical properties of codon context measures in an effort to better understand the phenomenon. We analyze the computational complexity of codon context optimization and design exact and efficient heuristic gene recoding algorithms under reasonable constraint models. We also present a web-based tool for evaluating codon context bias in the appropriate context.

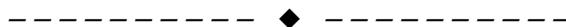

## 1 INTRODUCTION

Expression of genes is crucial for cell activity; it is a major determinant of phenotype as derived from genotype, and inherently fundamental to modern biotechnology. Expression is the process by which information from a gene is used in the synthesis of a functional gene product, most often a protein. Several steps in the gene expression process may be modulated, including the transcription, RNA splicing, translation and post-translational modification of a protein. We will concentrate on the process of translation, and the effect that synonymous mutations in a protein-coding gene confer to the expression of the corresponding protein, a method that has been traditionally employed to modify the post-transcriptional expression of genes [1]. Working towards the objectives of synthetic biology, precise protein expression control has direct implications in improving heterologous expression, and in successfully designing and fine-tuning gene regulatory networks.

Recent years have seen significant advances toward the understanding and control of the rate of translation of genes [1], [3], [4], [5], [6]. Newfound knowledge led to the development of a number of algorithms and computational tools that aim to enable life scientists to create their own synthetic genes and constructs [7]. A first generation of design tools focused primarily on optimizing designs for manufacturability (i.e. oligos without local secondary structures and end repeats) instead of biological activity. But soon the oligo design process was separated from the gene optimization process, and new tools emerged that address the two processes separately.

Codon bias, a characteristic pattern of preference in the usage of synonymous codons in highly and lowly expressed genes, has been shown to be correlated to expression levels of a gene and is a dominant method in designing synthetic genes for translational efficiency [6]. Codon context bias has emerged as another critical feature affecting gene translation rates independently of codon bias [8], but is less well studied and understood than the latter, and current methods and tools used to optimize codon context of a gene rely on heuristics that provide no guarantees of optimality.

In this study we examine statistical properties of codon context bias, in an effort to better understand the phenomenon and the measures used to quantify it. We also describe and analyze exact algorithms we designed for creating synthetic genes with optimized codon context, and evaluate the optimization potential of the simulated annealing heuristic as a method of choice toward efficient gene customization. Finally we make available a web-based utility to enable the characterization of a gene's codon pair bias, providing statistical information about its optimality.

## 2 BACKGROUND

### 2.1 Codon Bias

In most species, synonymous codons are used at unequal frequencies. Codon usage bias is recognized as crucial in shaping gene expression and cellular function, affecting diverse processes from RNA processing to protein translation and protein folding. Rarely used codons have been associated with rare tRNAs and have been shown to inhibit protein translation, where favorable codons have

- D. Papamichail is with the Department of Computer Science, The College of New Jersey, 2000 Pennington Road, Ewing, NJ 08628-0718, USA. E-mail: papamicd@tcnj.edu.
- H. Liu is with the Division of Biostatistics, Department of Public Health Science, Miller School of Medicine, University of Miami, Miami, FL 33136, USA.
- V. Machado and N. Gould are with the Department of Computer Science, The College of New Jersey, 2000 Pennington Road, Ewing, NJ 08628-0718, USA.
- J. R. Coleman is with the Department of Biology, Farmingdale State College (SUNY), Farmingdale, NY 11735-1021, USA.
- G. Papamichail is with the department of Informatics, New York College, Amalias 38 Ave., Athens, 10558, Greece.

the opposite effect, something that is particularly pronounced in prokaryotic organisms [9]. The process of substituting rare codons with favorable ones is referred to as codon optimization. Controlling codon bias without considering other optimization objectives, in order to modulate translation rates, is computationally easy, since it involves only certain synonymous substitutions to reach a desired distribution. The quantification of the effect though is much more difficult. Experimentally, the use of particular codons through synonymous mutations has been shown in certain cases to increase the expression of transgenes (genes expressed in a heterologous host) by more than 1000-fold [10].

Several different statistical methods have been proposed and used to analyze codon usage bias. Methods such as the *Frequency of optimal codons (Fop)* [11], the *Codon Adaptation Index (CAI)* [12], and the *tRNA adaptation index (tAI)* [13] are used to quantify codon preferences towards over- or under-represented codons, and to predict gene expression levels. Alternative methods such as the *Effective Number of codons (ENc)* and Shannon entropy from information theory [14] are used to measure codon usage evenness. *Relative Synonymous Codon Usage (RSCU)* [15] and *Synonymous Codon Usage Order (SCUO)* [16] are additional measures from the latter category.

Optimization of codon bias as a singular objective is algorithmically straightforward and can be performed in asymptotic linear time and space as a function of the sequence length, both in worst and average cases. This is true for maximization or minimization towards any given codon bias measure (such as CAI, RSCU, ENc, etc), as well as adoption/emulation of any given codon distribution, including the case when codon position assignments are performed randomly.

## 2.2 Codon Context Bias

Gutman and Hatfield first noticed that codon pairs in prokaryotic genes exhibit another significant bias towards specific combinations [17]. Further studies [18] revealed that codon pair optimization influences translational elongation step times, but their functional significance was studied only in very small datasets. More recent work by [8], [19], and [20] who synthesized novel coding regions utilizing large scale codon pair optimization and de-optimization, coupled with de novo synthesis of the constructs and in-vivo experimentation, provided further evidence of the influence codon pair bias has on translational efficiency.

Several mathematical methods have been proposed for the study of codon context bias, including [8], [21], [22], [23], [24], [25]. We are aware of at least three existing gene design tools that, as of date, provide functionality for controlling codon context [8], [26], [27].

*Computational Complexity*: Optimization of codon context as a singular objective has asymptotic linear time complexity in the worst case as a function of sequence length, as will be shown in subsection 4.1. Optimization of codon pair bias with a fixed codon distribution is considerably harder, although polynomially time solvable (section 4.2). This latter problem can be reduced to a vari-ation of the Travelling Salesperson Problem, which can be solved with a dynamic programming algorithm having $O(n^{42})$ time and $O(n^{41})$ space asymptotic complexity, where $n$ is the length of the sequence being optimized. As a consequence, all currently available tools that attempt to codon context optimize synthetic genes, often in conjunction with other objectives, utilize metaheuristics such as simulated annealing or genetic algorithms. These heuristics do not guarantee an optimal solution, but limit the running time of the optimization procedure, while typically computing reasonable approximations.

## 2.3 Codon Context Optimization Software

Several published gene design tools provide functionality for controlling codon context, albeit no two tools share the same measure of codon context bias. *Eugene* [26] is a standalone tool developed for multi-objective gene optimization. The program provides a graphical user interface and connects to databases such as GenBank to retrieve genomic sequences based on sequence identifiers. Among its numerous features, Eugene provides functionality to optimize mRNA codon context bias and codon autocorrelation. Eugene uses 'percentages' to indicate improvement towards a target objective instead of scores, which makes it difficult for one to interpret and compare to score-based results. The software provides two optimization methods, simulated annealing and a genetic algorithm, the former being significantly more efficient.

*Codon Optimization OnLine (COOL)* [27] is a web-based utility that can optimize for multiple objectives, including codon context bias. The optimization process uses a genetic algorithm to produce several approximately Pareto-optimal solutions given a set of design criteria. Randomly generated sequences are evaluated, ranked, and mutated until a stability threshold is reached, at which point the fittest sequences based on the chosen properties are outputted and the algorithm terminates. Codon context optimization is based on matching a given host codon pair distribution and no cumulative score is available to quantify the end result.

All current methods and tools have severe limitations, the most crucial being a lack of reference information about the optimization objectives and the optimality of the designs. Arbitrary scores are used to quantify codon and codon context bias, and it is hard to justify the use of one method over another. In the following sections we focus on codon context and study its statistical properties, as well as exact and approximate methods to evaluate the quality of an optimized design, which allow us to put codon context in the right context.

# 3 STATISTICAL PROPERTIES OF CODON CONTEXT BIAS

The phenomenon of unequal usage of synonymous codons is observed in most protein-coding genes, and is called codon bias [28]. A widely used measure that quantifies how far the codon usage of a gene departs from equal usage of synonymous codons is the *effective number of codons* used in a gene, $\widehat{N}_c$. This measure of synonymous codon usage bias is independent of the amino acid composition and the gene length [28].

$N_c$ [29] is calculated from the estimated homozygosity:

$$\hat{N}_c = 1/\hat{F}$$

where the estimated homozygosity $\hat{F}$ is:

$$\hat{F} = (n \sum_{i=1}^{k} p_i^2 - 1)/(n-1)$$

and $p_i$ is the frequency of usage of the synonymous codons obtained by dividing the actual usage $n_i$ by total usage of the amino acid $n$ (=$n_1 + \cdots + n_k$).

The $\hat{N}_c$ values for each of the 20 amino acids can be added together to get an *effective number of codons* $\hat{N}_c$ for the whole gene. Thus the value of $\hat{N}_c$ will vary from 20, when only one codon is used for each amino acid, to 61 when all codons are used equally for each amino acid.

*Codon Pair Bias* (*CPB*) refers to the phenomenon where synonymous codon pairs are used more or less frequently than expected [17]. For example, the amino acid pair Ala-Glu is expected to be encoded by GCCGAA and GCAGAG about equally frequently based on their codon frequencies. In fact, GCCGAA is used only one-seventh as frequently as GCAGAG in human genes ([8] supporting material).

A measure of *CPB* was defined in [8] from the following formula of *Codon Pair Score* (*CPS*) as follows:

$$CPS_{AB} = log \frac{O_{AB}}{E_{AB}} = log \frac{F(AB)}{\frac{F(A) * F(B)}{F(X) * F(Y)} * F(XY)} \quad (1)$$

where the codon pair *AB* encodes amino acid pair *XY*, *F* denotes frequency, *O* is the observed number of occurrences and *E* is the expected number of occurrences.

*CPS* of a given codon pair indicates whether the pair is over-represented (+) or under-represented (-) in a given genome.

Codon Pair Bias (*CPB*) for an entire gene sequence is the arithmetic mean of the codon pair scores of all pairs making up the entire gene sequence.

$$CPB_n = \sum_{i=1}^{k} \frac{CPS_i}{n} \quad (2)$$

where *n+1* is length of the gene sequence measured in codons.

Under the above definitions, the *CPS* scores for all 3721 possible codon pairs in human genes (excluding STOP codons) were calculated using the consensus CDS of the NCBI Reference Sequence Database, release date March 2nd 2005 ([8] supplementary material). *CPB* values for a main set of 14795 consistently annotated human genes were also calculated, and Figure 1B in [8] indicates a prevalent use of over-represent codon pairs in human genes. The mean codon pair bias for the 14795 annotated human genes is 0.07, which is also the peak of the distribution in that figure.

We examined statistical properties of Codon Pair Scores such as population mean and variance in the consistently annotated human protein coding gene set. In the analysis that follows we estimate the distribution of the *CPB* of human genes, based on which we calculate the *p-value* of a protein being encoded by an mRNA with a specific *CPB value*. We also demonstrate that codon pair bias is independent from codon bias and amino acid bias.

### 3.1 Distribution of Codon Pair Bias

First we will show that $CPS_{AB}$ and $CPS_{BC}$ are independent for different codon pairs *AB* and *BC*.

$$P(B|A) = \frac{P(AB)}{P(A)} = \frac{F(AB)/F(XY)}{F(A)/F(X)}$$

$$P(B) = F(B)/F(Y)$$

$$CPS_{AB} = log \frac{F(AB)}{\frac{F(A) * F(B)}{F(X) * F(Y)} * F(XY)} = log[\frac{P(B|A)}{P(B)}]$$

Similarly,

$$P(C|B) = \frac{P(BC)}{P(B)} = \frac{F(BC)/F(YZ)}{F(B)/F(Y)}$$

$$P(C) = F(C)/F(Z)$$

$$CPS_{BC} = log \frac{F(BC)}{\frac{F(B) * F(C)}{F(Y) * F(Z)} * F(YZ)} = log[\frac{P(C|B)}{P(C)}]$$

Since $P(B|A)$ and $P(C|B)$ are independent and P(B) and P(C) are approximately independent, $CPS_{AB}$ and $CPS_{BC}$ are independent.

Because $CPS_i$, *i*=1,2,…,*n*, are mutually independent each with the same distribution (though unknown), by the Central Limit Theorem,

$$CPB_n = \sum_{i=1}^{k} \frac{CPS_i}{n}$$

is approximately normally distributed for large n with mean E(*CPS*) and variance Var(*CPS*)/*n*, where *n+1* is the length of the gene sequence measured in codons.

By the definition of population mean and variance, E(*CPS*) is the weighted arithmetic mean of 3721 possible codon pairs (excluding the STOP codon) in the human ORFeome:

$$E(CPS) = \sum_{i=1}^{3721} [CPS_i \times \frac{N_{i,ob}}{N_{ob}}]$$

And Var(*CPS*) is calculated from

$$Var(CPS) = \sum_{i=1}^{3721} [(CPS_i - E(CPS))^2 \times \frac{N_{i,ob}}{N_{ob}}]$$

Based on the data found in the supplementary material of [8], E(*CPS*) in human genes is equal to 0.075 and Var(*CPS*) is 0.132. Thus $CPB_n$ is normally distributed with mean 0.075 and variance 0.132/n, results which match Figure 1B in [8] of CPBs calculated for a core set of 14795 consistently annotated human genes using the formulas of *CPS* and *CPB*.

The p-value of a protein being encoded by an mRNA with a specific codon pair bias *c* can be calculated as the two tail cumulative probability of the $CPB_n$:

$$p = P(CPB_n \leq c \ or \ CPB_n \geq 0.15 - c)$$
$$= 2 \int_{-\infty}^{c} \frac{1}{\sqrt{2\pi 0.132/n}} \exp(-\frac{(x - 0.075)^2}{2 * 0.132/n}) dx$$

Let 0.05 be the probability threshold of significance,

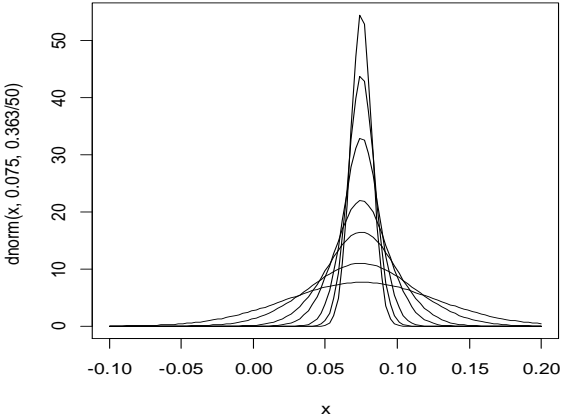

Figure 1: *CPB distribution of genes with length n = 49, 100, 225, 400, 900, 1600, 2500 (based on human ORFeome).*

where $p \leq 0.05$ implies that the probability of an $n+1$ length protein being encoded by an mRNA with a codon pair bias $c$ is rare. Then the 95% significant interval for $CPB_k$ is $(0.075-1.96*0.363/\sqrt{n}, \ 0.075+1.96*0.363/\sqrt{n})$. Reference examples:

For n=100, 95% significant interval for $CPB_{100}$ is (0.004, 0.146).

For n=400, 95% significant interval for $CPB_{400}$ is (0.039, 0.111).

For n=1600, 95% significant interval for $CPB_{1600}$ is (0.057, 0.093).

## 3.2 Independence of codon context bias from amino acid and codon bias

### 3.2.1 Discounting Amino acid pair bias
For randomly occurring amino acid pairs in a protein, the sum of the expected numbers of codon pairs encoding the same amino acid pair (400 amino acid pairs in total) should be equal to the sum of the observed numbers (1989, Gutman and Hatfield), i.e.

$$\sum_{kl} E_{kl} = \sum_{kl} O_{kl}$$

where $kl$ is any codon pair encoding the same amino acid pair $XY$.

For each group of codon pairs encoding the same amino acid pair $XY$, the expected values can be normalized when multiplied by a normalizing coefficient. This way, the sum of the expected values is the same as that of the observed values in each group.

$$E_{nor,ij} = \frac{\sum_{kl} O_{kl}}{\sum_{kl} E_{kl}} E_{ij}$$

Thus the normalized CPS of codon pair $AB$ is

$$CPS_{nor,AB} = log_{1.5} \frac{O_{AB}}{E_{nor,AB}}$$

Here we selected a log base to 1.5 such that the variance of $CPS$ is larger. $CPB_n$ can then vary in a larger range, reducing the effect of small numerical and rounding errors in resulting values.

### 3.2.2 Discounting codon bias
Expected occurrences of a given codon pair $AB$, $E_{AB}$, are calculated under the assumption that codons $A$ and $B$ are independent. If $A$ and $B$ are not independent, the ratio of $O_{AB}$ vs $E_{AB}$ will be much larger or smaller than 1, which indicates a nonrandom utilization of codon pairs. Since the nonrandom usage of codons is not related to the independence of $A$ and $B$, it cannot effect codon pair scores, thus does not contribute to the codon pair bias under the definition in this context.

## 4 ALGORITHMS FOR CODON CONTEXT OPTIMIZATION

In this section we will examine the problem of designing an mRNA encoding of a given protein that optimizes codon context bias. Optimization can occur either towards maximization or minimization of a given measure that quantifies codon context bias, such as codon pair bias as defined by Coleman et al. [8], codon context bias as defined by Moura et al. [25], the normalized offset value as defined by Boycheva et al. [22], and any other codon context bias measure that bounds the span (in nucleotides) of the context of a codon by a constant. In all subsequent discussion we will only consider the problem of bias maximization, where minimization can be achieved in the same exact manner.

We will concentrate on Codon Pair Score (*CPS*) and Codon Pair Bias (*CPB*), as defined in Equations (1) and (2) respectively, as the measures of choice to quantify codon context bias, because of the supporting experimental evidence for their validity [8], [19], [20], [30] ,[31] and our familiarity with their computation.

### 4.1 *Algorithm1*: Codon pair bias optimization without codon restrictions

We have designed and implemented an algorithm to compute the maximum and minimum *CPB* of a protein-coding gene in polynomial time using dynamic programming. Given a protein coding gene $G$ of length $n$, whose amino acid composition is $A_1 A_2 A_3 \ldots A_n$, we can calculate the mRNA sequence with the maximum (similar for minimum) *CPB* encoding the protein as follows:

*Algorithm1*
1. Starting with amino acid $A_1$, process each amino acid $A_i$, $1 \leq i \leq n$, in order, keeping track of the highest *CPB* of any mRNA prefix ending at each codon $C_{i,k}$, $1 \leq k \leq m_i$, where $m_i$ is the number of synonymous codons encoding amino acid $A_i$. Let $CPB_{i,k}$ be that highest score of any mRNA prefix ending with codon $C_{i,k}$ at position $i$. In addition, for each codon $C_{i,k}$, keep track of the *par-* *ent* codon (encoding amino acid $A_{i-1}$) that led to the highest *CPB* prefix score ending at that codon.

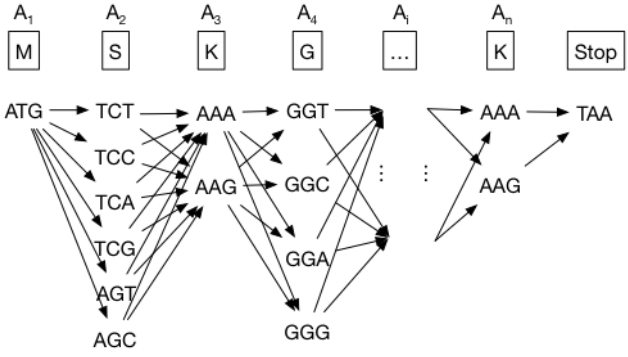

Figure 2: Dynamic programming algorithm for codon context optimization

2. At each step of the algorithm, when proceeding from amino acid $A_i$ to $A_{i+1}$, form all possible codon pairs encoding the amino acid pair $A_iA_{i+1}$. Compute the maximum *CPB* of all mRNA prefixes ending at each codon $C_{i+1,l}$, $1 \le l \le m_{i+1}$, as $max(CPB_{i,k} + CPS(C_{i,k}, C_{i+1,l}))$, for $1 \le k \le m_i$, then store that value at codon $C_{i+1,l}$, together with the appropriate parent codon. Here the notation $CPS(A, B)$ is used instead of $CPS_{AB}$ for clarity.

3. When all amino acids are processed, select the codon with the highest *CPB* score among all synonymous codons of $A_n$. That value is the highest *CPB* of any mRNA encoding of gene $G$. The mRNA encoding with the highest *CPB* can be generated by following *parent* pointers toward the start of the amino acid chain and concatenating the appropriate codons.

An outline of the data structure used by *Algorithm1* can be observed in Figure 2.

**Theorem 1.** *Algorithm1 computes the optimal CPB of a protein coding gene in O(n) worst case time and space complexity, where n is the length of amino acid sequence of the gene.*

**Proof.** For algorithm correctness it is sufficient to show proposition $P_1(n)$ = "*Algorithm1* identifies the mRNA with the highest *CPB* translating into a given protein $A_1A_2...A_n$, for integer $n \ge 2$." Proposition $P_1(n)$ follows from a strengthened proposition $P_2(n)$ which is easier to prove using induction: $P_2(n)$ = "*Algorithm1* identifies the mRNA with the highest *CPB* translating into a given protein $A_1A_2...A_n$, $n \ge 2$, and ending at codon $C_{n,k}$, $1 \le k \le m_n$", where $m_n$ is the number of synonymous codons translating to amino acid $A_n$.

*Base Case:* After the first step of the algorithm, the codon pair with the highest *CPS* encoding $A_1A_2$ and ending at codon $C_{2,k}$, $1 \le k \le m_2$ can be formed following the parent pointer to concatenate the two codons. The codon pair formed will have the highest *CPB* among all synonymous ones, since only the highest *CPB* value is stored at each codon at position 2. Therefore $P_2(2)$ is true.

*Inductive Step:* Assuming that proposition $P_2(q)$ is true

for an arbitrary but fixed integer $q$ such that $2 \le q \le n-1$, we will show that $P_2(q+1)$ is true. We claim that on its $q$-th step *Algorithm1* will compute and store at codon $C_{q+1,k}$ the highest *CPB* score of any mRNA encoding of the prefix of the protein ending at position $q+1$ and having $C_{q+1,k}$ as its last codon. This is true since all possible codon pairs $C_{q,l}C_{q+1,k}$ are considered, for $1 \le l \le m_q$, and are concatenated with the optimal mRNA encodings ending at codons $C_{q,l}$, according to our inductive hypothesis. As such, selecting the mRNA sequence ending at codon $C_{q+1,k}$ with the highest *CPB* score among these, and storing that score and the appropriate parent pointer at codon $C_{q+1,k}$, allows the identification of the mRNA sequence with the optimal *CPB* ending at codon $C_{q+1,k}$.

Therefore $P_2(n)$ is true for all integers $n \ge 2$, from which it follows that $P_1(n)$ is also true for all $n \ge 2$.

The linear time and space complexities of *Algorithm1* follow from the linear number of steps that the algorithm takes, at each of which a constant number of codon pairs is processed (at most 36, since the maximum number of synonymous codons for any amino acid is 6) and a constant amount of information is stored, namely the *CPB* values for each prefix ending at a given synonymous codon, and the parent pointers, both quantities bounded by the constant 6. □

### 4.2 Algorithm2: Polynomial *CPB* Optimization with Fixed Codon Distribution

Due to the significant effect of codon bias in mRNA translation rates, it is important to control codon bias while recoding an amino acid sequence by optimizing its codon pair bias. For that reason we concentrate on optimizing *CPB* while assigning a fixed codon distribution to our mRNA sequence, which remains unchanged during our *CPB* optimization procedure.

The problem of maximizing the presence of over- or under-represented codon pairs in an mRNA sequence, while keeping the amino acid sequence and codon frequency distributions intact, can be reduced to a variant of the TSP (Travelling Salesperson Problem). This particular variant, which can be thought as computing the shortest route to travel a series of countries (amino acids) in a pre-specified sequence, and selecting the city (codon) we visit in each country from a given fixed distribution of cities, is polynomially solvable using dynamic programming with a time complexity of $O(n^{42})$, where $n$ is the number of codons in the examined mRNA sequence. We will now describe another algorithm, *Algorithm2*, which polynomially solves the aforementioned problem.

The algorithm is similar to *Algorithm1* for *CPB* optimization without codon restrictions. In this particular case though we have to keep track not only of the optimal *CPB* of mRNA prefixes ending at synonymous codons at each amino acid position, but also of all allocated codons in the prefixes, as future codon choices may be limited based on the selections made so far. We use dynamic programming to reduce the exponential number of codon permutations, utilizing the fact that knowledge of specific codon assignments at each position is not necessary, as only the total number of codons of each type that have been already allocated restrict further selections. So it is sufficient to keep track of codon frequencies and the optimal *CPB* for every prefix ending at a specific codon having used a specific codon distribution. Since any codon may be placed in the mRNA of the protein a maximum of $n$ times, and we have 61 non-terminating codons, at each amino acid position we have to keep track of a maximum of $6 \times n^{61}$ values. The algorithm executes for a total of $n$ steps, resulting in a worst-case time complexity of $O(n^{62})$, with worst-case space utilization of $O(n^{61})$.

The time complexity of the algorithm can be reduced to $O(n^{42})$ by considering the synonymous codon degrees of freedom. For every amino acid encoded by $k$ synonymous codons we only need to keep track of the frequencies of $k$-1 of its codons, since the frequency of the last codon can be derived from the rest. The improved time bound is therefore a result of the 41 degrees of freedom of codons. Similarly, the space required by this algorithm is $O(n^{41})$. Even so, this algorithm, despite being polynomial, is not practical except for the smallest of protein coding genes.

The correctness of this algorithm can be argued in a similar manner with the proof of Theorem 1.

### 4.3 Algorithm3: Branch and Bound *CPB* Optimization with Fixed Codon Distribution

To study the effectiveness of heuristics in optimizing *CPB* of synthetic genes with fixed codon distributions, we need a reference algorithm that produces the optimal encodings against which to compare approximate solutions. We designed a branch and bound algorithm to calculate the maximum (and minimum) *CPB* of a given gene with a given codon distribution which examines all possible mRNA encodings of the gene with fixed codon frequencies.

The algorithm simply enumerates all possible encodings, by assigning to each amino acid (in order) any corresponding codon from the pool of available codons. It performs a depth first search on the space of codon assignments at each position of the amino acid chain, where the explored tree has a height of $n$ and each internal node has a maximum of 6 children. This process is depicted in Figure 3.

To eliminate branches leading to provably sub-optimal solutions, we compare the best gene *CPB* result thus far with the potential of the current prefix *CPB* when added to the optimal *CPB* of the suffix of the mRNA, as determined by *Algorithm1* which considers any possible codon assignments. Under *CPB* maximization, the initial bound can be an arbitrarily negative number or a previously computed sub-optimal design with a method such as simulated annealing (described in section 4.4).

We pre-compute the optimal *CPB* of all suffixes of our mRNA by running the dynamic programming *Algorithm1* starting at the end of the amino acid chain, and storing all intermediate results in an array. This provides constant lookup time for the optimal *CPB* under any codon distribution of any suffix of the given gene. The worst-case asymptotic time complexity of *Algorithm3* is exponential as a function of protein length, which remains exponen-tial in practice, as demonstrated in section 5. The worst-case space complexity is linear owing to the depth first search process and constant amount of information stored at each step of the algorithm, in addition to the linear-size precomputed suffix *CPB* upper bound scores.

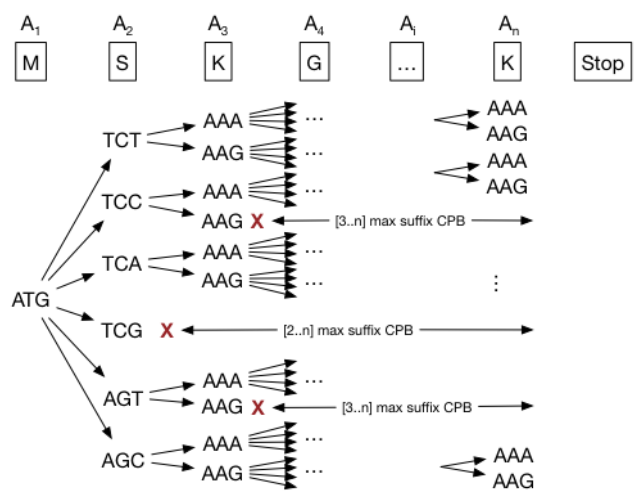

Figure 3: Branch and bound CPB optimization algorithm with fixed codon distribution

### 4.4 Algorithm4: Simulated Annealing for CPB Optimization with Fixed Codon Distribution

Simulated annealing [32] is a fast-converging metaheuristic well suited for TSP-like problems [33], [34]. It is particularly useful for problems such as synthetic gene design, since it can be effectively applied to optimize an mRNA sequence for multiple objectives in addition to codon pair bias, such as restriction site inclusion/elimination, and RNA folding energy manipulation.

We have used single-objective simulated annealing extensively in the past to design the synthetic viral and bacterial genes that were featured in [8], [19], and [20], and have experimented with (conflicting) multi-objective simulated annealing in optimizing polio virus capsid protein encodings for codon pair utilization and CpG content (unpublished data), where the codon pair bias effect on translation was determined to be independent from the CpG content of coding sequences, at least for the RNA encoding of the capsid protein of Polio virus.

Our simulated annealing procedure works as follows:
*Algorithm4*
1. We initially create a random assignment of the codons in their respective amino acid allowed positions. If a new codon distribution is defined, first we assign codons to each amino acid according to the new distribution and then create the random assignment.
2. We calculate the codon pair bias of the coding sequence from the initial assignment.
3. We randomly select an amino acid and then randomly consider two locations where this amino acid is encoded in our coding sequence. If the codons at these positions are different, we calculate the sum of scores of the four newly formed codon pairs as if these codons were exchanged, and compare it to the sum of the original four pairs. If there is an improvement to the overall score, we exchange the two codons. If not, we calculate the simulated annealing optimization function and decide whether to exchange the two codons based on its value. This way, even non-beneficial exchanges have a well-defined probability of being performed, with this probability varying throughout the execution based on the simulated annealing parameters and algorithm progress.

4. We repeat the previous step until we have reached the maximum number of allowed steps, which is specified at the execution of the algorithm.

## 5 EXPERIMENTAL RESULTS

One of the goals of our study is to determine whether the simulated annealing heuristic produces *CPB* designs that are sufficiently close to optimal, as to reduce the need to run computationally expensive algorithms for further optimization. Toward that goal we implemented two algorithms to maximize *CPB* of protein coding genes and compare their results, *Algorithm3*, and *Algorithm4*, the former producing the maximal *CPB* design, and the latter an approximation. We also implemented *Algorithm1* to aid the branch and bound algorithm in generating the necessary upper bounds on mRNA suffixes' *CPB* scores.

We run all algorithmic experiments on a laptop equipped with a Haswell (22nm) i7-4900mq processor running at 2.8GHz, 32GB of memory and an SSD hard drive. All algorithms run on a single processor core.

We produced maximal and near-maximal *CPB* designs for prefixes of the Aequorea victoria Green Fluorescent Protein (GFP), GenBank accession number M62653.1, that ranged in size from 10 to 70 amino acids. The upper size limit was set by computational time constraints of *Algorithm3*, which grows exponentially as a function of the amino acid sequence length. *Algorithm4*, which utilizes simulated annealing, was run for 500,000 iterations in each computational experiment, independent of protein size. The algorithm was run 5 times for each input protein prefix, and the best result was collected, although the results for each input length were always the same and optimal. The running time of *Algorithm4* is dependent primarily on the number of iterations, and for all experiments presented herein ranged between 3.7 and 4.6 seconds. The *CPB* outcomes of the two algorithms and actual running time of *Algorithm3* are shown in Table 1.

### TABLE 1
CPB OPTIMIZATION ALGORITHM RESULTS

| Sequence length (AAs) | *Algorithm3* running time (seconds) | Maximal *CPB* (*Algorithms* 3 & 4) |
|---|---|---|
| 10 | 0.086 | 0.294582188343425 |
| 20 | 0.065 | 0.298431996808867 |
| 30 | 0.100 | 0.298816298402260 |
| 40 | 0.790 | 0.268955938539740 |
| 50 | 9.501 | 0.246521370447020 |
| 60 | 734.195 | 0.242171188081159 |
| 70 | 327337.980 | 0.233378389895599 |

The simulated annealing procedure (*Algorithm4*), whose parameters have been customized experimentally in the past to produce optimized designs of the Poliovirus P1 capsid protein [8] and numerous other genes from a variety of organisms, manages to match the optimal score of *Algorithm3* for all GFP prefixes tested, up to 70 amino acids long. Although it is expected that differences will emerge as protein size increases, it is reassuring that, despite the already sizable solution space explored by the tested cases, simulated annealing identifies the optimal solution. Our belief that simulated annealing is a great choice for *CPB* optimization is corroborated by p-values of *CPB* scores for optimized mRNA encodings of larger proteins, which are consistently placed a large number of standard deviations from the respective mean.

Figure 4: Codon Context tool (CCtool) input screen

## 6 CODON CONTEXT EVALUATION TOOL

We have created a web-based tool that can compute the *CPB* of a given gene and graphically place the score in a range of values determined by the minimum and maximum *CPB* of the gene when approximated by simulated annealing. We also randomized codon assignments to corresponding amino acids and generated 100 alternate designs coding for the same protein, having the exact same codon distribution, to approximate the mean and variance of the CBP of that gene. An example of the input and output screen of our tool are shown in Figures 4 and 5 respectively. The tool is still under development and can be accessed online at http://algo.tcnj.edu/cctool.

## 6 CONCLUSION

Codon context bias optimization significantly affects mRNA translation rates and is currently employed in several tools that design and evaluate synthetic protein coding genes. Due to the computational complexity of optimizing codon context, current methods utilize heuristics that provide little to no information on solution optimality. In addition, no two methods employ the same measure of codon context, and results are not comparable.

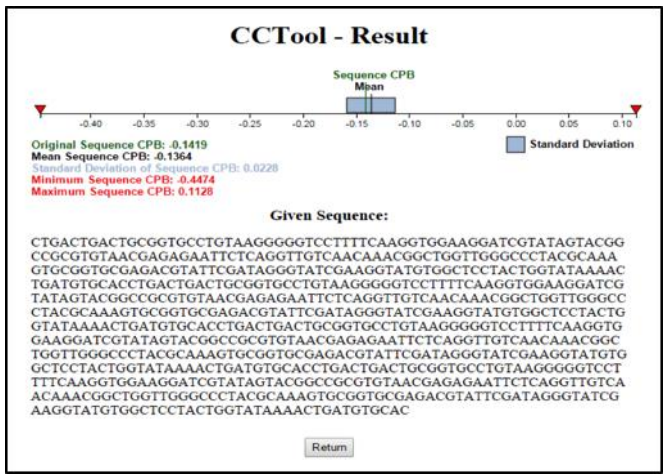

Figure 5: CCtool output page.

We performed a statistical analysis to better understand codon pair bias, the only experimentally validated codon context measure, and determined its independence from amino acid and codon biases. We also designed exact algorithms to optimize mRNA sequences for codon pair bias under different constraints, and run experiments to determine the performance of simulated annealing for codon context optimization. Finally we developed a web-based utility to enable synthetic gene designers to evaluate their own constructs and put codon context in the right context.

## ACKNOWLEDGMENT

This work has been supported by NSF Grant CCF-1418874 and the MUSE program at The College of New Jersey. D. Papamichail is the corresponding author and can be reached at papamicd@tcnj.edu.